\documentclass[12pt]{iopart}

\usepackage{graphicx}

\begin{document}

\title[\underline{\footnotesize\rm Interpolation formula for the reflection coefficient distribution 
of absorbing chaotic cavities}]
{Interpolation formula for the reflection coefficient distribution 
of absorbing chaotic cavities in the presence of time reversal symmetry}

\author{M Mart\'{\i}nez-Mares$^1$ and R A M\'endez-S\'anchez$^2$}
\address{$^1$ Departamento de F\'{\i}sica, Universidad Aut\'onoma
Metropolitana-Iztapalapa, 09340 M\'exico D. F., M\'exico}
\address{$^2$ Centro de Ciencias F\'{\i}sicas, Universidad Nacional Aut\'onoma de
M\'exico, A.P. 48-3, 62210, Cuernavaca, Mor., M\'exico}

\date{\today}

\begin{abstract}
We propose an interpolation formula for the distribution of the 
reflection coefficient in the presence of time reversal symmetry 
for chaotic cavities with absorption. This is done assuming a 
similar functional form as that when time reversal invariance is 
absent. The interpolation formula reduces to the analytical 
expressions for the strong and weak absorption limits. Our proposal is compared to the quite complicated exact result existing in the literature. 
\end{abstract}





\section{Introduction}
\label{sec:intro}

In recent years there has been a great interest in the study of 
absorption effects on transport properties of classically chaotic cavities \cite{Doron1990,Lewenkopf1992,Brouwer1997,Kogan2000,Beenakker2001,Schanze2001,
Schafer2003,Savin2003,Mendez-Sanchez2003,Fyodorov2003,Fyodorov2004,Savin2004,
FyodorovSavin2004,Hemmady2004,Schanze2005,Kuhl2005,Savin2005,MMM2005} (for a review 
see Ref.~\cite{Fyorev}). This is due to the fact that for experiments in microwave cavities~\cite{Richter,Stoeckmann}, elastic resonators~\cite{Schaadt} and elastic media~\cite{Morales2001} absorption always is present. Although the external parameters are particularly easy to control, absorption, due to power loss in the volume of the device used in the experiments, is an ingredient that has to be taken into account in the verification of the 
random matrix theory (RMT) predictions. 

In a microwave experiment of a ballistic chaotic cavity connected to 
a waveguide supporting one propagating mode, Doron 
{\it et al}~\cite{Doron1990} studied the effect of absorption on the 
$1\times 1$ sub-unitary scattering matrix $S$, parametrized as 
\begin{equation} 
S=\sqrt{R}\, e^{i\theta}, 
\label{S11}
\end{equation}
where $R$ is the reflection coefficient and $\theta$ is twice the 
phase shift. 
The experimental results were explained by Lewenkopf 
{\it et al.}~\cite{Lewenkopf1992} by simulating the absorption 
in terms of $N_p$ 
equivalent ``parasitic channels", not directly accessible to experiment, 
each one having an imperfect coupling to the cavity described by the 
transmission coefficient $T_p$.

A simple model to describe chaotic scattering including absorption 
was proposed by Kogan {\it et al.}~\cite{Kogan2000}. It describes 
the system through a sub-unitary scattering matrix $S$, whose 
statistical distribution satisfies a maximum information-entropy 
criterion. Unfortunately the model turns out to be valid only in 
the strong-absorption limit and for $R\ll 1$. 
For the $1\times 1$ $S$-matrix of 
Eq.~(\ref{S11}), it was shown that in this limit $\theta$ is 
uniformly distributed between 0 and $2\pi$, while $R$ satisfies 
Rayleigh's distribution
\begin{equation}
P_{\beta}(R) = \alpha e^{-\alpha R}; \qquad R \ll 1, 
\hbox{ and } \alpha \gg 1, 
\label{Rayleigh}
\end{equation}
where $\beta$ denotes the universality class of $S$ introduced by 
Dyson~\cite{Dyson}: $\beta=1$ when time reversal invariance (TRI) is 
present (also called the {\it orthogonal} case), $\beta=2$ when TRI 
is broken ({\it unitary} case) and $\beta=4$ corresponds to the 
symplectic case. 
Here, $\alpha=\gamma\beta/2$ and $\gamma=2\pi/\tau_a\Delta$, is the 
ratio of the mean dwell time inside the cavity ($2\pi/\Delta$), where 
$\Delta$ is the mean level spacing, and $\tau_a$ is the 
absorption time. 
This ratio is a measure of the 
absorption strength. 
Eq.~(\ref{Rayleigh}) is valid for $\gamma\gg 1$ and for $R \ll 1$ 
as we shall see below. 

The weak absorption limit ($\gamma\ll 1$) of $P_{\beta}(R)$ was 
calculated by Beenakker and Brouwer~\cite{Beenakker2001}, by relating 
$R$ to the time-delay in a chaotic cavity which is distributed according 
to the Laguerre ensemble. The distribution of the reflexion coefficient 
in this case is  
\begin{equation}
P_{\beta}(R) = \frac{\alpha^{1+\beta/2}}{\Gamma(1+\beta/2)} 
\frac{e^{-\alpha/(1-R)}}{(1-R)^{2+\beta/2}}; \qquad \alpha\ll 1.
\label{Laguerre}
\end{equation} 
In the whole range of $\gamma$, $P_{\beta}(R)$ 
was explicitly obtained for $\beta=2$~\cite{Beenakker2001}:
\begin{equation} 
P_2(R) = \frac{e^{-\gamma/(1-R)}}{(1-R)^3}
\left[ \gamma (e^{\gamma}-1) + (1+\gamma-e^{\gamma}) (1-R) \right],
\label{beta2}
\end{equation}
and for $\beta=4$ more recently~\cite{FyodorovSavin2004}.
Eq.~(\ref{beta2}) reduces to Eq.~(\ref{Laguerre}) for small 
absorption ($\gamma\ll 1$) while for strong absorption it becomes 
\begin{equation} \label{bigbeta2}
P_2(R) = \frac{\gamma \, e^{-\gamma R/(1-R)}}{(1-R)^3}; 
\qquad \gamma\gg 1. 
\end{equation}
Notice that $P_2(R)$ approaches zero for $R$ close to one. 
Then the Rayleigh distribution, Eq.~(\ref{Rayleigh}), 
is only reproduced in the range of few standard deviations 
i.e., for $R \stackrel{<}{\sim} \gamma^{-1}$. This can be 
seen in Fig.~\ref{fig:fig1}(a) where we compare the distribution 
$P_2(R)$ given by Eqs.~(\ref{Rayleigh}) and~(\ref{bigbeta2}) with the 
exact result given by Eq.~(\ref{beta2}) for $\gamma=20$. 
As can be seen the result obtained from the time-delay agrees with 
the exact result but the Rayleigh distribution is only valid for 
$R\ll 1$. 

Since the majority of the experiments with absorption are performed with 
TRI ($\beta=1$) it is very important to have the result in this case. 
Due to the lack of an exact expression at that time, 
Savin and Sommers~\cite{Savin2003} proposed an approximate 
distribution $P_{\beta=1}(R)$ by replacing $\gamma$ by $\gamma\beta/2$ in 
Eq.~(\ref{beta2}). However, this is valid for the intermediate and strong 
absorption limits only. Another formula was proposed in 
Ref.~\cite{Kuhl2005} as an interpolation between the strong and 
weak absorption limits assuming a quite similar expression as the 
$\beta=2$ case (see also Ref.~\cite{FyodorovSavin2004}). 
More recently~\cite{Savin2005}, a formula for the integrated 
probability distribution of $x=(1+R)/(1-R)$, 
$W(x)=\int_x^\infty P_0^{(\beta=1)}(x)dx$, was obtained. The 
distribution 
$P_{\beta=1}(R)=\frac 2{(1-R)^2}P_0^{(\beta=1)}(\frac{1+R}{1-R})$ 
then yields a quite complicated formula. 

Due to the importance to have an ``easy to use'' formula for 
the time reversal case, our purpose is to propose a better 
interpolation formula for $P_{\beta}(R)$ when $\beta=1$. In the next 
section we do this following the same procedure as in 
Ref.~\cite{Kuhl2005}. 
We verify later on that our proposal reaches both limits 
of strong and weak absorption. In Sec.~\ref{sec:conclusions} we compare 
our interpolation formula with the exact result of Ref.~\cite{Savin2005}. 
A brief conclusion follows. 

\section{An interpolation formula for $\beta=1$}

From Eqs.~(\ref{Rayleigh}) and~(\ref{Laguerre}) we note that $\gamma$ 
enters in $P_{\beta}(R)$ always in the combination 
$\gamma\beta/2$. We take this into account and combine it with the 
general form of $P_2(R)$ and the interpolation proposed in 
Ref.~\cite{Kuhl2005}. For $\beta=1$ we then propose the following formula 
for the $R$-distribution 
\begin{equation}
P_1(R) = C_1(\alpha) 
\frac{ e^{-\alpha/(1-R)} }{ (1-R)^{5/2} }
\left[ \alpha^{1/2} (e^{\alpha}-1) + 
(1+\alpha-e^{\alpha}) 
{}_2F_1 \left( \frac 12,\frac 12,1;R \right)\frac{1-R}2 \right], 
\label{beta1}
\end{equation}  
where $\alpha=\gamma/2$, ${}_2F_1$ is a hyper-geometric 
function~\cite{Abramowitz}, and $C_1(\alpha)$ is a normalization 
constant
\begin{equation}
C_1(\alpha) = \frac{\alpha}
{ (e^{\alpha} - 1) \Gamma(3/2,\alpha) + 
\alpha^{1/2}( 1 + \alpha - e^{\alpha} ) f(\alpha)/2 }
\end{equation}
where 
\begin{equation} 
f(\alpha) = \int_{\alpha}^{\infty} \frac{e^{-x}}{x^{1/2}} \, \, {}_2F_1 
\left( \frac 12,\frac 12,1;1-\frac{\alpha}{x}\right)
\end{equation}
and $\Gamma(a,x)$ is the incomplete $\Gamma$-function
\begin{equation}
\Gamma(a,x) = \int_x^{\infty} e^{-t} t^{a-1} dt.
\label{Gammafunc} 
\end{equation}

In the next sections, we verify that in the limits of strong and 
weak absorption we reproduce Eqs.~(\ref{Rayleigh}) and~(\ref{Laguerre}).


\section{Strong absorption limit}

\begin{figure}
\begin{center}
\includegraphics[width=8.0cm]{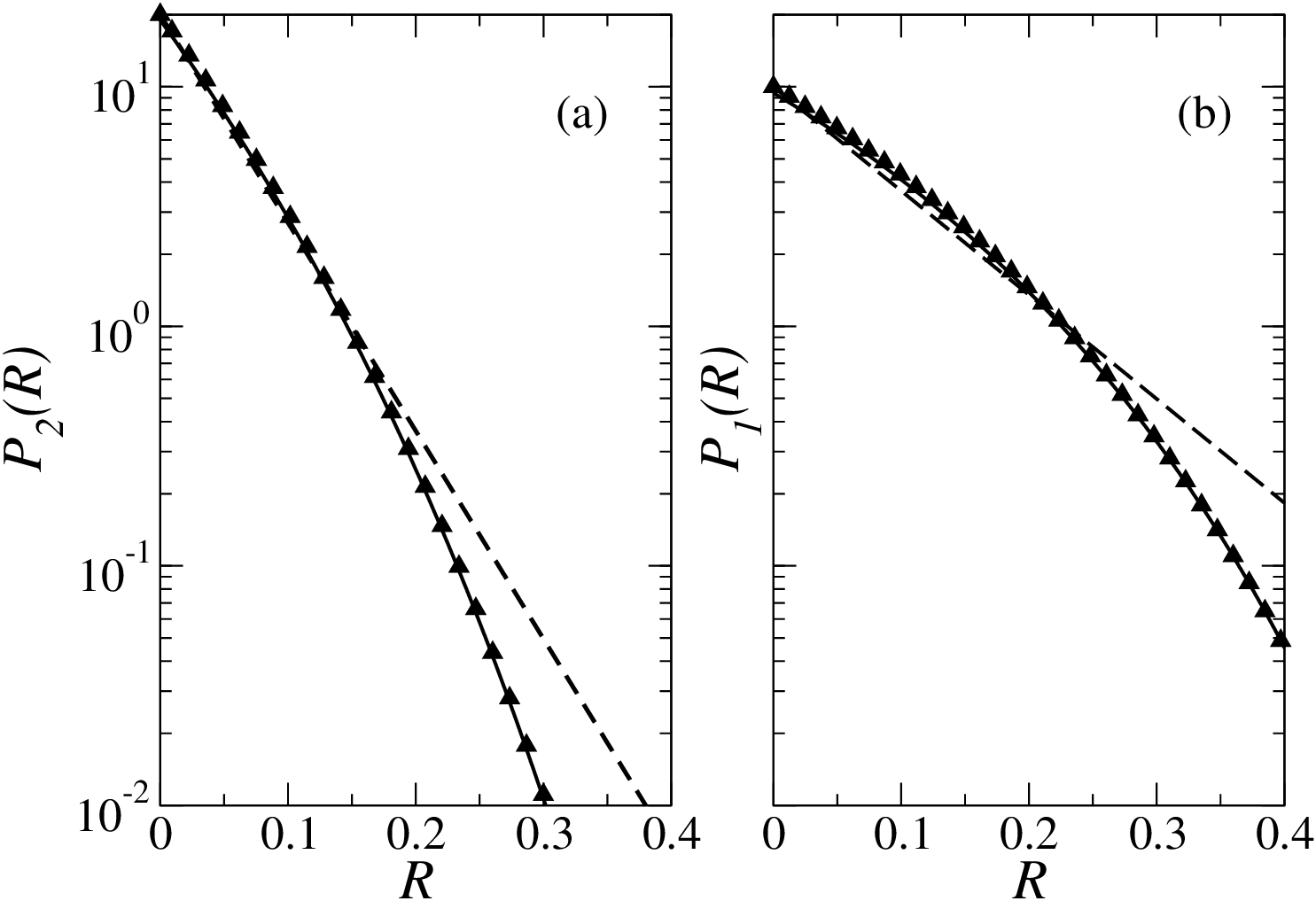}
\caption{Distribution of the reflection coefficient for absorption 
strength  $\gamma=20$, for (a) $\beta=2$ (unitary case) and (b) 
$\beta=1$ (orthogonal case). 
In (a) the continuous line is the exact result Eq.~(\ref{beta2}) while 
in (b) it corresponds to the interpolation formula, 
Eq.~(\ref{beta1}). 
The triangles in (a) are the results given by Eq.~(\ref{bigbeta2}) for 
$\beta=2$ and in (b) they correspond to Eq.~(\ref{bigbeta1}).
The dashed line is the Rayleigh distribution Eq.~(\ref{Rayleigh}), 
valid only for $R\stackrel{<}{\sim}\gamma^{-1}$ and $\gamma\gg1$.
}
\label{fig:fig1}
\end{center}
\end{figure}

In the strong absorption limit, $\alpha\rightarrow\infty$, 
$\Gamma(3/2,\alpha)\rightarrow\alpha^{1/2}e^{-\alpha}$, and 
$f(\alpha)\rightarrow\alpha^{-1/2}e^{-\alpha}$. Then, 
\begin{equation}
\lim_{\alpha\rightarrow\infty} C_1(\alpha) = 
\frac{\alpha e^{\alpha}}{ (e^{\alpha}-1)\alpha^{1/2} + 
(1+\alpha-e^{\alpha})/2} \simeq 
\alpha^{1/2}.
\end{equation} 
Therefore, the $R$-distribution in this limit reduces to  
\begin{equation}\label{bigbeta1}
P_1(R) \simeq \frac{ \alpha \, e^{-\alpha R/(1-R)} }{ (1-R)^{5/2} } 
\qquad \alpha \gg 1 ,
\end{equation}
which is the equivalent of Eq.~(\ref{bigbeta2}) but now for $\beta=1$. 
As for the $\beta=2$ symmetry, 
it is consistent with the fact that $P_1(R)$ approaches zero as $R$ 
tends to one. It reproduces also Eq.~(\ref{Rayleigh}) in the range 
of a few standard deviations ($R\stackrel{<}{\sim}\gamma^{-1}\ll 1$), as can be seen 
in Fig.~\ref{fig:fig1}(b).


\section{Weak absorption limit}

For weak absorption $\alpha\rightarrow 0$, the incomplete 
$\Gamma$-function 
in $C_1(\alpha)$ reduces to a $\Gamma$-function $\Gamma(x)$ 
[see Eq.~(\ref{Gammafunc})]. Then, $P_1(R)$ can be written as 
\begin{eqnarray} 
 P_1(R) && \simeq \frac{\alpha}
{ (\alpha+\alpha^2/2+\cdots)\Gamma(3/2)-
(\alpha^{5/2}/2+\cdots )f(0)/2 } \nonumber \\ 
&& \times 
\frac{ e^{-\alpha/(1-R)} }{(1-R)^{5/2}} 
\big[ \alpha^{3/2} + \alpha^{5/2}/2 +\cdots \nonumber \\ 
&& - ( \alpha^2/2 + \alpha^3/6 +\cdots){}_2F_1(1/2,1/2,1;R)(1-R)/2 \big] .
\end{eqnarray}
By keeping the dominant term for small $\alpha$, Eq.~(\ref{Laguerre}) 
is reproduced.


\section{Comparison with the exact result}

\begin{figure}
\begin{center}
\includegraphics[width=8.0cm]{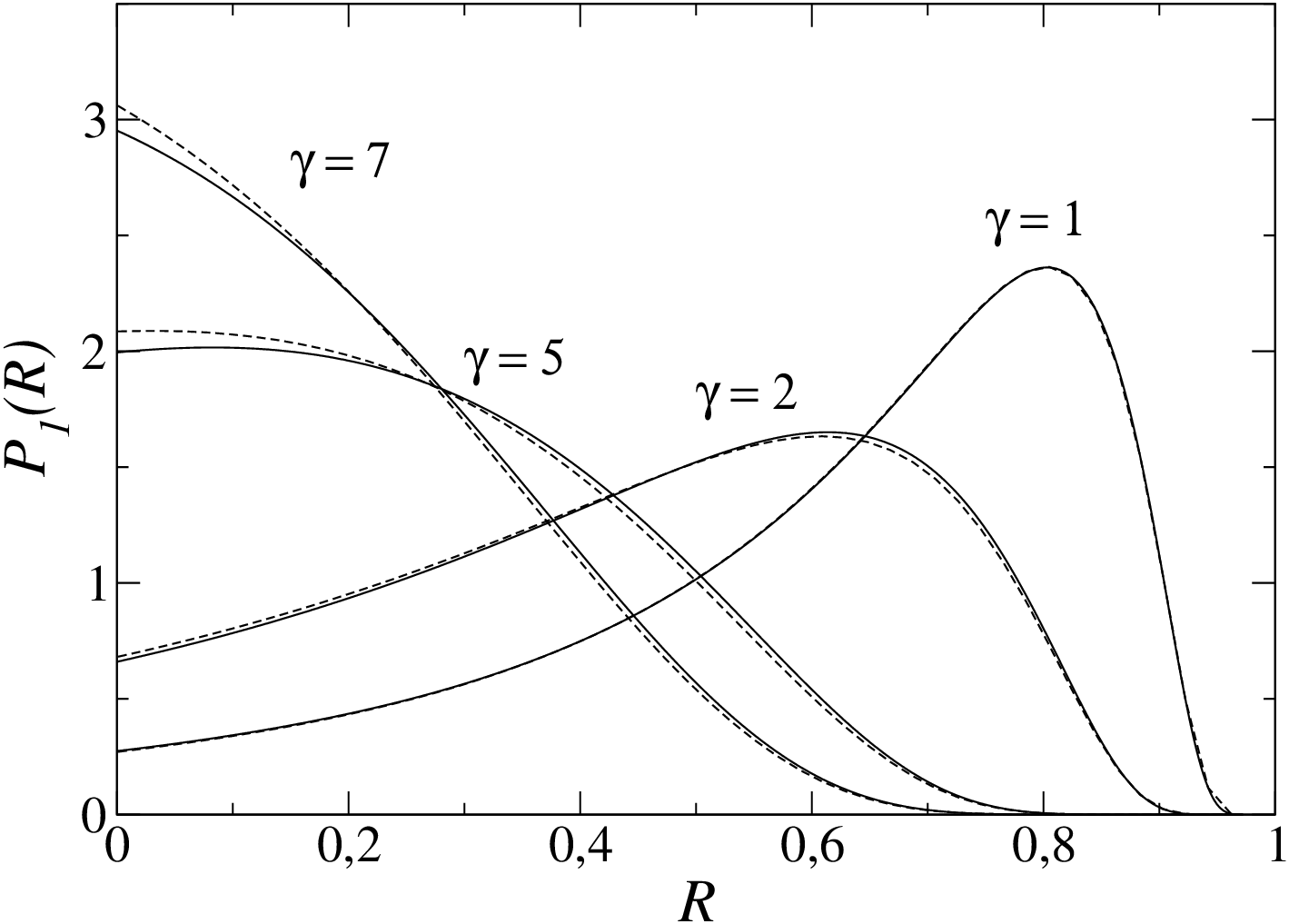}
\caption{Distribution of the reflection coefficient in the presence of 
time-reversal symmetry for absorption strength $\gamma=1$, 2, 5, and 7. 
The continuous lines correspond to the distribution given by 
Eq.~(\ref{beta1}). For comparison we include the exact results 
of Ref.~\cite{Savin2005} (dashed lines).}
\label{fig:fig2}
\end{center}
\end{figure}

\begin{figure}[b]
\begin{center}
\includegraphics[width=8.0cm]{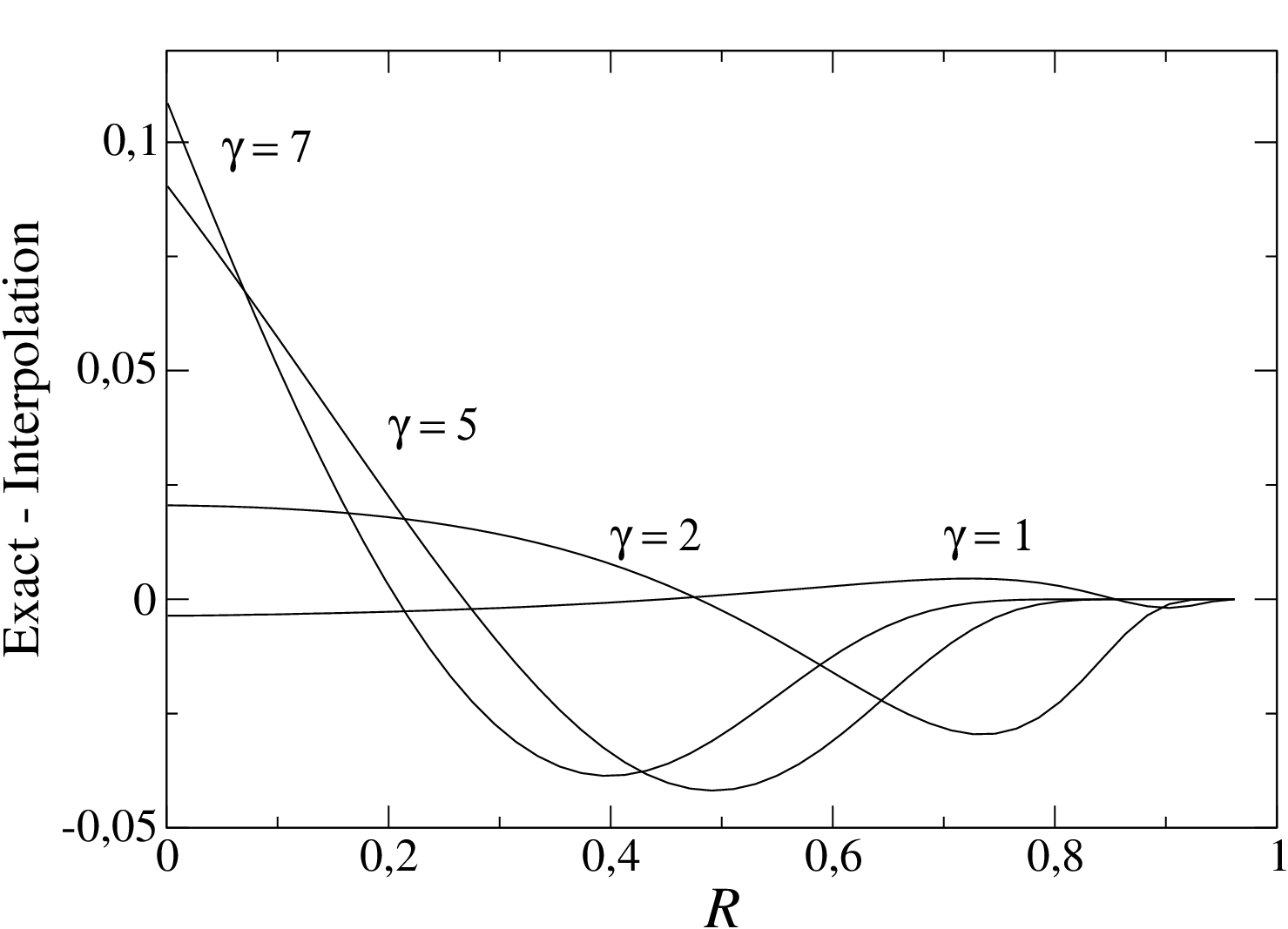}
\caption{Difference between the exact result and the interpolation 
formula, Eq.~(\ref{beta1}), for the $R$-distribution for $\beta=1$ 
for the same values of $\gamma$ as in Fig.~\ref{fig:fig2}.}
\label{fig:fig3}
\end{center}
\end{figure}

In Fig.~\ref{fig:fig2} we compare our interpolation formula, 
Eq.~(\ref{beta1}), with the exact result of Ref.~\cite{Savin2005}. 
For the same parameters used in that reference we observe an 
excellent agreement. 
In Fig.~\ref{fig:fig3} we plot the difference between 
the exact and the interpolation formulas for the same values 
of $\gamma$ as in Fig.~\ref{fig:fig2}. The error 
of the interpolation formula is less than 4\%.

\section{Conclusions}
\label{sec:conclusions}

We have introduced a new interpolation formula for the 
reflection coefficient 
distribution $P_{\beta}(R)$ in the presence of time reversal symmetry 
for chaotic cavities with absorption. The interpolation formula 
reduces to the analytical expressions for the strong and weak 
absorption limits. Our proposal is to produce an ``easy to use'' 
formula that differs by a few percent from the exact, but 
quite complicated, result of Ref.~\cite{Savin2005}. 
We can summarize the results for both symmetries ($\beta=1$, 2) 
as follows 
\begin{equation}
P_{\beta}(R) = C_{\beta}(\alpha) 
\frac{ e^{-\alpha/(1-R)} }{ (1-R)^{2+\beta/2} }
\left[ \alpha^{\beta/2} (e^{\alpha}-1) + 
(1+\alpha-e^{\alpha}) 
{}_2F_1 \left(\frac{\beta}2,\frac{\beta}2,1;R\right) 
\frac{\beta(1-R)^{\beta}}2 \right], 
\end{equation}  
where $C_{\beta}(\alpha)$ is a normalization constant that depends on 
$\alpha=\gamma\beta/2$. This interpolation formula is exact 
for $\beta=2$ and yields the correct limits of strong and weak 
absorption.

\ack

The authors thank to DGAPA-UNAM for financial support through 
project IN118805. We thank D. V. Savin for provide us the data 
for the exact results we used in Figs.~\ref{fig:fig2} and~\ref{fig:fig3} 
and to J. Flores and P. A. Mello for useful comments.



\end{document}